\documentclass[aps,prl,twocolumn,superscriptaddress]{revtex4-2}
\usepackage{amssymb}
\usepackage{amsmath,amsthm}
\usepackage[dvipsnames]{xcolor}
\usepackage{chngpage}
\usepackage{lgrind}        
\usepackage{color}         
\usepackage{graphics}      
\usepackage[pdftex]{graphicx}      
\usepackage{longtable}     
\usepackage{epsf}          
\usepackage{bm}            
\usepackage{asymptote}     
\usepackage{thumbpdf}
\usepackage{verbatim}
\usepackage{url}
\usepackage{MnSymbol}
\usepackage{physics}
\usepackage[colorlinks=true]{hyperref} 
\usepackage{enumitem}	

\usepackage{float}
\usepackage{times}
\usepackage{color}
\usepackage{verbatim}
\usepackage{soul,xcolor}
\setstcolor{red}
\usepackage{colortbl}
\usepackage[normalem]{ulem} 

\newcommand{\cut}[1]{} 
\newcommand{\outline}[1]{}   

\usepackage{tabularx}
\usepackage{bbold}

\makeatletter
\AtBeginDocument{
\g@addto@macro{\appendix}{\renewcommand{\p@subsection}{\@Alph\c@section}}
}
\makeatother

\begin{document}

\title{Dressed-state Hamiltonian engineering in a strongly interacting solid-state spin ensemble}

\affiliation{Department of Physics, Harvard University, Cambridge, MA 02138, USA}
\affiliation{Materials Department, University of California, Santa Barbara, CA 93106, USA}
\affiliation{Division of Engineering and Applied Science, California Institute of Technology, Pasadena, CA 91125, USA}
\affiliation{Harvard Quantum Initiative, Harvard University, Cambridge, MA 02138, USA}
\affiliation{Department of Chemistry and Chemical Biology, Harvard University, Cambridge, MA 02138, USA}
\affiliation{Department of Physics, University of California, Santa Barbara, CA 93106, USA}

\author{Haoyang Gao$^1$}
\author{Nathaniel T. Leitao$^1$}
\author{Siddharth Dandavate$^1$}
\author{Lillian B. Hughes Wyatt$^{2,3}$}
\author{Piotr Put$^1$}
\author{Mathew Mammen$^4$}
\author{Leigh S. Martin$^1$}
\author{Hongkun Park$^{1,5}$}
\author{Ania C. Bleszynski Jayich$^6$}
\author{Mikhail D. Lukin$^{1,\dagger}$}

\begin{abstract}
In quantum science applications, ranging from many-body physics to  quantum metrology,
dipolar interactions in spin ensembles are controlled via Floquet engineering. However, this technique typically reduces the interaction strength between spins, and effectively weakens the coupling to a target sensing field, limiting the metrological sensitivity. In this work, we develop and demonstrate a method for direct tuning of the native interaction in an ensemble of nitrogen-vacancy (NV) centers in diamond. Our approach utilizes dressed-state qubit encoding under a magnetic field perpendicular to the crystal lattice orientation. This method leads to a $3.2\times$ enhancement of the dimensionless coherence parameter $JT_2$ compared to state-of-the-art Floquet engineering, and a $2.6\times$ ($8.3$~dB) enhanced sensitivity in AC magnetometry. Utilizing the extended coherence we experimentally probe spin transport at intermediate to late times. 
Our results provide a powerful Hamiltonian engineering tool for future studies with NV ensembles and other interacting higher-spin ($S>\frac{1}{2}$) systems.
\end{abstract}
\maketitle

The robust coherent control of strongly interacting quantum systems is an essential tool in many areas of quantum science.
In systems with fixed native interactions—such as ensembles of nitrogen-vacancy (NV) centers in diamond—Floquet engineering\cite{Choi2020Floquet} has become the standard approach for tuning interactions, achieved through weighted averaging over toggling frames connected by microwave (MW) pulse sequences. This method has established NV ensembles as a promising platform for nanoscale magnetic sensing\cite{Schirhagl2014, Casola2018, Mamin2013, Hong2013} and quantum many-body physics\cite{Choi2017TimeCrystal,Kucsko2018,Martin2023}, combining flexible tunability with inherent advantages such as large system sizes, room-temperature operation\cite{Balasubramanian2009}, and ease of integration with sensing targets\cite{Ofori2012}.

However, such tunability comes at a cost. First, the weighted averaging process inherent to Floquet engineering weakens the effective interaction strength $J$ between spins, as the opposite-signed Ising and exchange parts of the native Hamiltonian\cite{Kucsko2018} (i.e. $H_{ij}\propto s_i^x s_j^x + s_i^y s_j^y - s_i^z s_j^z$ for a pair of spins $i$ and $j$) partially cancel each other during  the averaging procedure. As a direct consequence, the dimensionless coherence parameter $JT_2$ is often limited to a relatively small value  ($\sim$3--4, see e.g. Ref.~\cite{Martin2023}). This often prevents the exploration of  
 interesting many-body phenomena - such as emergent hydrodynamics\cite{Zu2021,Peng2023,Stasiuk2025} - that occur at intermediate to late timescales. Similarly, in quantum sensing applications, such averaging inevitably transforms the target field into a weaker effective field\cite{Zhou2020}, limiting the achievable sensitivity.

\begin{figure}
\begin{center}
\includegraphics[width=\columnwidth]{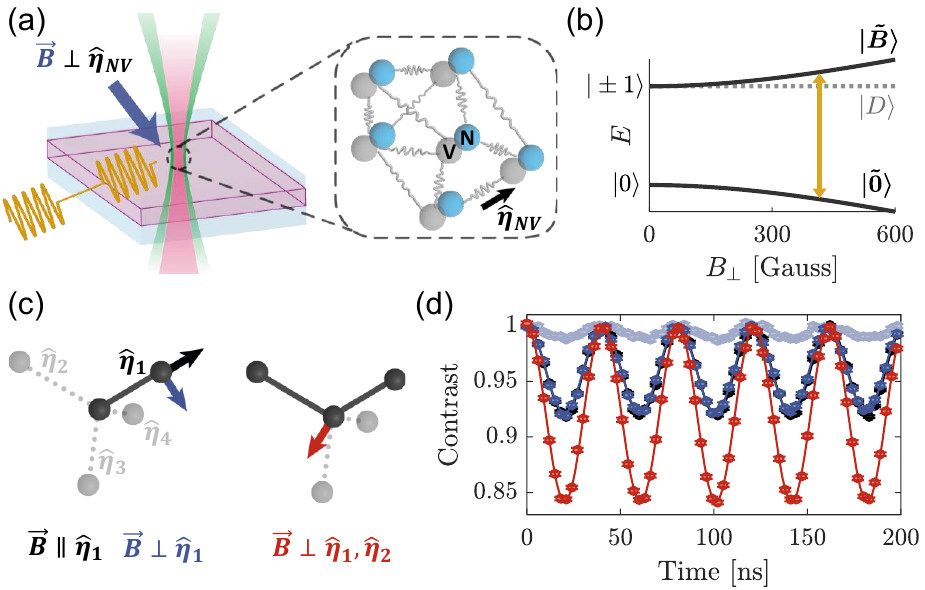}
\caption{{\bf{Overview of the experiments}.} (a-b) The experiments are performed in an interacting ensemble of NV centers\cite{Hughes2023} polarized optically by green laser within a confocal spot of a microscope. An external B-field perpendicular to the NV crystal lattice orientation is applied, and the qubit is encoded in the resulting dressed-states $\ket{\tilde{0}}$ and $\ket{\tilde{B}}$. Global microwave drive is used to manipulate the spins, and the spin-state of the ensemble is readout via the red fluorescence intensity collected with confocal microscopy. (c) The three configurations used in this work, including on-axis field (black), field perpendicular to one group of NVs (blue), and field simultaneously perpendicular to two groups of NVs (red). The same color label is used throughout the paper. (d) Comparison of Rabi oscillation under the three configurations. The semi-transparent blue trace represents the perpendicular field configuration without the pulsed field (see Ref.~\cite{Gao2025} for pulsed field), and the solid blue trace represents the case with the pulsed field, which restores high quality readout. The beat node free oscillation in the two-groups configuration (red) indicates the same Rabi frequency shared by the two groups.}
\label{fig1}
\end{center}
\end{figure}

In this Letter, we present direct tuning of the native dipolar interaction between NV centers, through the application of a magnetic field perpendicular to the NV axis $\hat{\eta}_{\mathrm{NV}}$ (Fig.~\ref{fig1}(a)). 
Although field-tunable dipolar interaction is not a new concept and has been exploited in platforms such as polar molecules\cite{Li2023}, it has not been used for NV centers or other solid-state spin ensembles, as experiments have primarily focused on B-fields parallel to the NV axis (i.e. the on-axis field configuration), where such tunability is impossible\cite{Kucsko2018}. The use of the off-axis field is associated with two challenges. First, the optical spin contrast decreases rapidly as a function of the off-axis field\cite{Tetienne2012}, which disrupts NV's native initialization and readout mechanisms\cite{Doherty2013}. 
In addition, off-axis field configurations are typically accompanied by non-commuting couplings between the NV electronic spin and its host nitrogen nuclear spin, leading to complicated entangling dynamics such as electron-spin-echo-envelope-modulation (ESEEM) effect\cite{Rowan1965} and faster decay of the electronic spin coherence.

In this work,  we overcome the first challenge utilizing a pulsed B-field introduced in Ref.~\cite{Gao2025}, which adiabatically switches between the off-axis configuration for spin dynamics and the on-axis configuration for initialization and readout. As seen in Fig.~\ref{fig1}(d), this technique nearly perfectly restores the spin contrast (blue and black traces). To overcome the second challenge, we note that although a generic off-axis field leads to ESEEM effect, it can be avoided under a B-field exactly perpendicular to the NV axis\cite{Qiu2021}. 

\begin{figure}
\begin{center}
\includegraphics[width=\columnwidth]{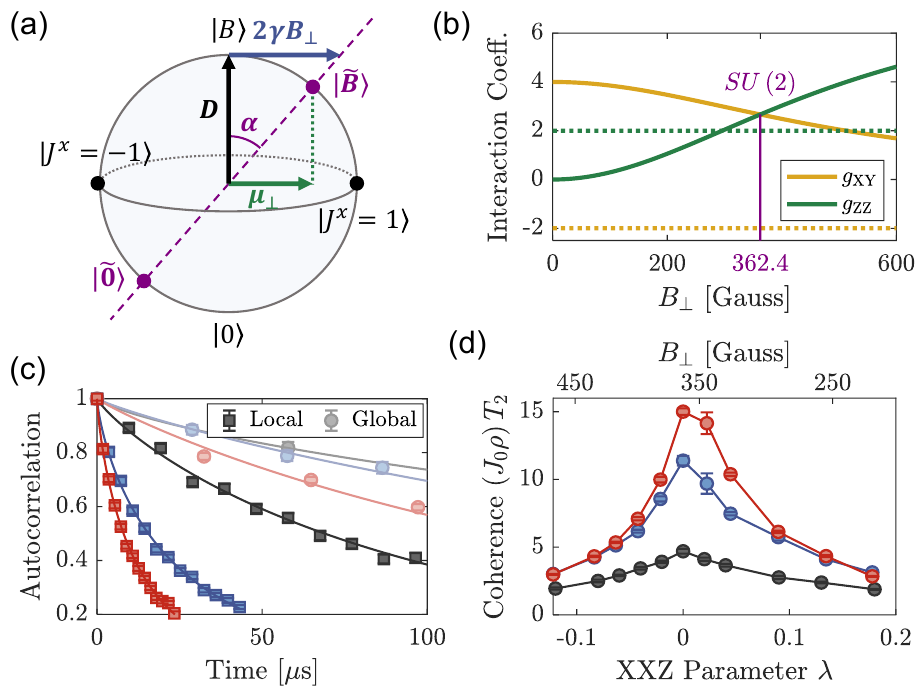}
\caption{{\bf{Tunability of native interaction}.} (a) Bloch sphere illustration of the dressed-states encoding on $\ket{\tilde{0}}$ and $\ket{\tilde{B}}$. The factor 2 in the effective field $2\gamma B_\perp$ comes from quantum interference enhancement of the coupling between $\ket{0}$ and $\ket{B}\equiv\left(\ket{+1}+\ket{-1}\right)/\sqrt{2}$. As $B_\perp$ increases, the dressed-states obtains larger permanent magnetic moment ($\mu_\perp$, green), leading to stronger Ising coupling. (b) Theoretically predicted exchange (yellow) and Ising (green) parts of the interaction, featuring a transition from easy-plane to easy-axis Hamiltonian as $B_\perp$ increase. A native Hamiltonian with $\mathrm{SU}\left(2\right)$ symmetry is expected at $B_\perp = 362.4$~Gauss. (c) Disorder-order measurement\cite{Martin2023} of the infinite temperature local auto-correlator $C_{\mathrm{Local}}^{\mathrm{XX}}\equiv\overline{\langle s_i^x\left(t\right)s_i^x\left(0\right)\rangle_{T=\infty}}$ (solid square markers), in comparison with the decay of global X polarization (semi-transparent circle markers) at the $\mathrm{SU}\left(2\right)$ point. Colors indicates the three different configurations in Fig.~\ref{fig1}(c). The $\mathrm{SU}\left(2\right)$ symmetric Hamiltonian is Floquet engineered in the on-axis field configuration (black), and is native in the other two cases. The faster decay of $C_{\mathrm{Local}}^{\mathrm{XX}}$ in the perpendicular field configurations indicates an enhanced interaction strength. The reported data are normalized against extrinsic decay, similar to Ref.~\cite{Martin2023}. See Supplement for raw data and effects of such normalization. (d) Comparison of the dimensionless coherence parameter $\left(J_0\rho\right) T_2$ for the three configurations, where $\rho$ is the independently characterized NV density (see Supplement). The XXZ anisotropy $\lambda$ is tuned via $B_\perp$ in the perpendicular field configurations (upper X axis), 
and via Floquet engineering in the on-axis field case (see Supplement for details on Floquet pulse sequences).
}
\label{fig2}
\end{center}
\end{figure}

\begin{figure}
\begin{center}
\includegraphics[width=\columnwidth]{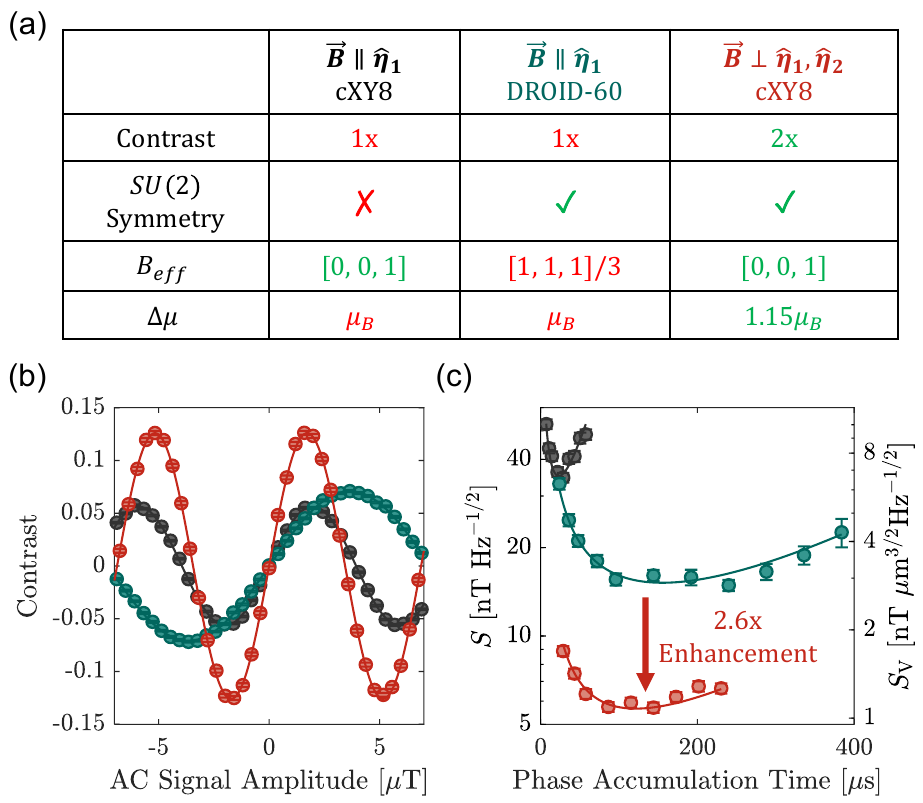}
\caption{{\bf{Demonstration of sensitivity enhancements}.} (a) 
Comparison of the pros (green) and cons (red) of the three sensing scenarios: on-axis field configuration with and without Floquet engineering (DROID-60), and the two-groups perpendicular field configuration at native $\mathrm{SU}\left(2\right)$ point. The pulse sequence ``cXY8" stands for XY8 concatenated with itself\cite{Khodjasteh2005}, which is a modified version of XY8 that has better robustness against coherent pulse errors (see Supplement). (b) Comparison of the sensing signal response in the three scenarios, while fixing the phase accumulation time to be 7.2~$\mu$s in all three scenarios. The reported AC signal amplitude (i.e. the horizontal axis) is the component along the respective direction of magnetic moment $\Delta\vec{\mu}$ (i.e. along the original NV axis for the on-axis field configuration, and along $\hat{B}_\perp$ for the perpendicular field configuration), for fair comparison between them. (c) Comparison of the sensitivity between the three scenarios. The solid lines indicate the theoretical sensitivity scaling expected from the measured coherence time. The volume normalized sensitivity is calculated with an estimated confocal spot diameter 500~nm and NV layer thickness 185~nm. 
}
\label{fig3}
\end{center}
\end{figure}

\begin{figure}
\begin{center}
\includegraphics[width=\columnwidth]{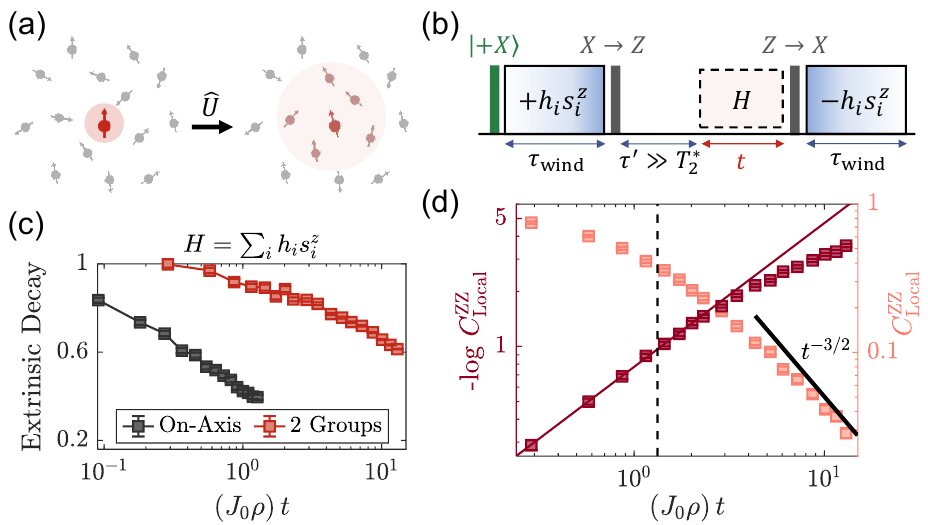}
\caption{{\bf{Spin transport measurements}.} (a) Illustration of spin transport in an infinite temperature background, where the initial polarization of the central spin spreads to a larger region under unitary evolution. (b) The modified disorder-order\cite{Martin2023} protocol for measuring local auto-correlator $C_{\mathrm{Local}}^{\mathrm{ZZ}}\equiv\overline{\langle s_i^z\left(t\right)s_i^z\left(0\right)\rangle_{T=\infty}}$. The disorder winding time is chosen as $\tau_\mathrm{wind}\sim 3.5T_2^\star$, and the dephasing time is chosen as $\tau^\prime = 3\tau_\mathrm{wind}$ (see Supplement for discussions). (c) Extrinsic decay in the above protocol due to imperfect disorder unwinding, for the on-axis field configuration (black) and two-groups perpendicular field configuration (red). Operationally, this data is measured under free evolution (i.e. no MW pulses in the dashed box ``$H$" in panel (b)), where the strong on-site disorder $\sum_i h_i s_i^z$ freezes the transport. (d) Measured $C_{\mathrm{Local}}^{\mathrm{ZZ}}\left(t\right)$ at the $\mathrm{SU}\left(2\right)$ point in the two-groups perpendicular field configuration, with on-site disorder decoupled via concatenated XY8 sequence\cite{Khodjasteh2005} (see Supplement). The data is normalized against extrinsic decay (see Supplement for raw data). The dark red line is an early-time stretched exponential fit, with fitting range including all data points where $C_{\mathrm{Local}}^{\mathrm{ZZ}}\geq 0.25$. The solid black line represents $t^{-3/2}$ decay, for comparison purposes. The dashed vertical line indicates extrinsic decay timescale in the on-axis field configuration (panel (c), black).}
\label{fig4}
\end{center}
\end{figure}

Working in such a perpendicular field configuration, we demonstrate a tunable interaction with a $3.2\times$ enhanced dimensionless coherence parameter $JT_2$ (Fig.~\ref{fig2}) compared to Floquet engineering. 
We then apply our technique to AC magnetometry by tuning the native interaction to the $\mathrm{SU}\left(2\right)$ point that preserves the spin coherence, and realize a $2.6\times$ (i.e. $8.3$~dB) enhancement of the sensitivity (Fig.~\ref{fig3}) compared to Floquet-based approaches\cite{Zhou2020}. 
Finally, we demonstrate a unique application of our approach in studying intermediate to late time spin transport (Fig.~\ref{fig4}), as enabled by the extended coherence and the highly static on-site disorder (Fig.~\ref{fig4}(c)) in the perpendicular field configuration.

\textit{Dressed-state engineering of dipolar interaction.}---The ground-state spin Hamiltonian of a single NV center in a perpendicular field $B_\perp$  along x-axis is
\begin{equation}
    H_{\mathrm{NV}} = D\left(J^z\right)^2 + \gamma B_\perp J^x,
\label{eq:NV_external_field}
\end{equation}
where $\vec{J}$ is the spin-1 operator of the NV electronic spin, $D = 2.87~\mathrm{GHz}$ is the zero-field splitting, and $\gamma = 2.8~\mathrm{MHz/G}$ is the gyromagnetic ratio. To understand the spectrum (Fig.~\ref{fig1}(b)) under such Hamiltonian, we notice that $J^x$ couples $\ket{0}$ to the bright state $\ket{B}\equiv\left(\ket{+1}+\ket{-1}\right)/\sqrt{2}$, while leaving the dark state $\ket{D}\equiv\left(\ket{+1}-\ket{-1}\right)/\sqrt{2}$ as an eigenstate decoupled from the perpendicular field. Therefore, the system simplifies to a two-level system spanned by $\left\{\ket{0}, \ket{B}\right\}$, whose eigenstates are represented geometrically in Fig.~\ref{fig2}(a):
\begin{align}
    \ket{\tilde{0}} &= \cos{\frac{\alpha}{2}}\ket{0}-\sin{\frac{\alpha}{2}}\ket{B} \nonumber\\
    \ket{\tilde{B}} &= \cos{\frac{\alpha}{2}}\ket{B}+\sin{\frac{\alpha}{2}}\ket{0},
\end{align}
where $\alpha=\tan^{-1}\left(\frac{2\gamma B_\perp}{D}\right)$ is the mixing angle of the dressed-states.

As shown in Fig.~\ref{fig1}(b), we encode the qubit on the $\ket{\tilde{0}}$ and $\ket{\tilde{B}}$ levels, where the effective interaction between the qubits can be calculated by projecting the full dipolar Hamiltonian 
\begin{equation}
    H_{ij}^{\mathrm{Dipole}} = - \frac{J_{\mathrm{Dipole}}}{r^3}\left[3\left(\vec{J_i}\cdot\hat{r}\right)\left(\vec{J_j}\cdot\hat{r}\right) - \vec{J_i}\cdot\vec{J_j}\right]
\label{eq: original_dipolar_interaction}
\end{equation}
onto the qubit subspace, with $J_{\mathrm{Dipole}} \equiv \left(2\pi\right)~52~\mathrm{MHz}\cdot\mathrm{nm}^3$ and $\vec{r}$ being the distance between the $i^\mathrm{th}$ and $j^\mathrm{th}$ spin. An important observation is that the projection of $J^y$ and $J^z$ are zero
\begin{equation}
    \mathcal{P}\left(J^y\right) = \mathcal{P}\left(J^z\right) = 0,
\label{eq:only_Jx_coupling}
\end{equation}
as both $\ket{\tilde{0}}$ and $\ket{\tilde{B}}$ are even under the transformation $\ket{m}\leftrightarrow\ket{-m}$, but $J^y$ and $J^z$ are odd. As a direct consequence, two NV centers only interact through their magnetic moments along $\hat{x}$ (or equivalently, $\hat{B}_\perp$), which naturally becomes the effective quantization axis that sets the dipolar anisotropy. This leads to a generic form of the effective interaction:
\begin{equation}
    H_{ij}^{\mathrm{eff}} = -\frac{J_{\mathrm{Dipole}}}{r^3}A_{\hat{B}_\perp}\left(\hat{r}\right)\left[g_{\mathrm{XY}}\left(s_i^x s_j^x + s_i^y s_j^y\right) + g_{\mathrm{ZZ}}s_i^z s_j^z\right],
\label{eq:effective_interaction}
\end{equation}
where $\vec{s}$ is the effective spin-$\frac{1}{2}$ operator defined in the qubit subspace, $A_{\hat{B}_\perp}\left(\hat{r}\right)\equiv\frac{3\left(\hat{B}_\perp\cdot\hat{r}\right)^2-1}{2}$ is the dipolar anisotropy, and $g_{XY}$, $g_{ZZ}$ are the exchange and Ising coefficients of the Hamiltonian.

The dependence of $g_{XY}$ and $g_{ZZ}$ on $B_\perp$ can be calculated by projecting $J^x$ onto the qubit subspace (see Supplement\cite{SI}), which leads to:
\begin{equation}
    g_{\mathrm{XY}} =4 \cos^2\alpha,~~~~~~~~g_{\mathrm{ZZ}} =8 \sin^2\alpha,
\end{equation}
as plotted in Fig.~\ref{fig2}(b). Here we see that the form of the Hamiltonian can be tuned from the easy-plane side to the easy-axis side by increasing $B_\perp$, as intuitively expected from the increasing permanent magnetic moments of the eigenstates (Fig.~\ref{fig2}(a), $\mu_\perp$), which is the origin of the Ising coupling. In addition, the strength of the interaction is enhanced by a factor of 4 compared to the on-axis field configuration (dashed lines in Fig.~\ref{fig2}(b)), as measured by the Hamiltonian trace $2g_{\mathrm{XY}}+g_{\mathrm{ZZ}}$. Such measure is chosen as an invariant under Floquet engineering\cite{Choi2020Floquet}, offering a fair comparison between Hamiltonians with different XXZ anisotropy, assuming a Floquet engineering that tunes them into the same form. The factor of 4 enhancement can be understood intuitively in the strong $B_\perp$ limit, where the qubit is encoded between $\ket{J^x=+1}$ and $\ket{J^x = -1}$, and the doubled dipole moments lead to the quadrupled interaction.

To further enhance the interaction strength, one can bring two groups of NV centers with different crystal lattice orientation into resonance by aligning the B-field to be simultaneously perpendicular to them (Fig.~\ref{fig1}(c)), doubling the effective density of interacting spins. A concern for this field configuration is that the inter-group interaction may be complicated, and the two groups may experience different coupling to the external MW driving. However, this is not a case, as Eq.~(\ref{eq:only_Jx_coupling}) guarantees that the two groups interact with each other and with the MW field through the identical component of their magnetic moment (i.e. the $\hat{x}$, or equivalently, $\hat{B}_\perp$ component), leading to a scenario where the inter- and intra- group interactions are exactly the same, in addition to the identical coupling of the two groups to the MW driving. This expectation is confirmed in Fig.~\ref{fig1}(d), where the beat node free Rabi oscillation demonstrates the same control shared by the two groups.

\textit{Observation of tunable interaction.}---We now turn to experimental observation of our predictions. The first experiment we did is to test the enhanced interaction strength at the native $\mathrm{SU}\left(2\right)$ point. Although this form of interactions preserves the total spin polarization and is not detectable through global measurements, it contributes to decay of the infinite temperature local auto-correlator $C_{\mathrm{Local}}^{\mathrm{XX}}\equiv\overline{\langle s_i^x\left(t\right)s_i^x\left(0\right)\rangle_{T=\infty}}$ that can be detected through the ``disorder-order" measurement introduced in Ref. \cite{Martin2023}. In Fig.~\ref{fig2}(c), we observe 
the expected 4x faster decay in the perpendicular field configuration (blue) compared to the Floquet engineered $\mathrm{SU}\left(2\right)$ Hamiltonian in the on-axis field case (black), and a further 2x enhancement in the two-groups configuration (red). We further compare the decay of local auto-correlator to the decay of global X polarization (Fig.~\ref{fig2}(c), semi-transparent  traces), which is conserved under the $\mathrm{SU}\left(2\right)$ Hamiltonian and thus purely reflects extrinsic decoherence.
Although the coherence time measured in absolute units is shorter in the perpendicular field configurations, such effect is not as significant as the enhancement of interaction strength, which leads to an overall enhancement in the dimensionless coherence parameter $JT_2$.

We further verify the tunability of the native Hamiltonian by measuring the global decay timescale $T_2$ of an initial state polarized along the $\mathrm{X}$ direction, as a function of $B_\perp$ (Fig.~\ref{fig2}(d)). Here, we reparametrize the interaction (Eq.~(\ref{eq:effective_interaction})) as
\begin{equation}
    H_{ij}^{\mathrm{eff}} = -\frac{J_0}{r^3}A_{\hat{B}_\perp}\left(\hat{r}\right)\left[\vec{s}_i\cdot\vec{s}_j + \lambda\left(s_i^x s_j^x + s_i^y s_j^y - 2s_i^z s_j^z\right)\right]
\end{equation}
to separate the XXZ anisotropy 
$\lambda\equiv \frac{g_\mathrm{XY}-g_\mathrm{ZZ}}{2g_\mathrm{XY}+g_\mathrm{ZZ}}$ 
from the overall strength 
$J_0\equiv J_\mathrm{Dipole} \frac{2g_\mathrm{XY}+g_\mathrm{ZZ}}{3}$, 
 which is convenient for comparison with Floquet engineered Hamiltonian in the on-axis field configuration.
Fig.~\ref{fig2}(d) shows a sharp peak whose location is consistent with the theoretically expected $\mathrm{SU}\left(2\right)$ point, providing direct experimental evidence on the tunability of the Hamiltonian. In addition, the dimensionless coherence parameter $JT_2\equiv\left(J_0\rho\right)T_2$ - with $\rho$ being the NV density - exceeds 15 in the two-groups perpendicular field configuration (red), enabling exploration of many-body physics at intermediate to late timescales. 

\textit{Applications in quantum sensing.}---The ability to engineer a Hamiltonian with $\mathrm{SU}\left(2\right)$ symmetry is crucial for applications in AC magnetometry, as the coherence time is no longer limited by spin-spin interactions. Comparing with existing methods based on Floquet engineering, such as the ``DROID-60" pulse sequence\cite{Zhou2020}, a key advantage of the \textit{native} $\mathrm{SU}\left(2\right)$ symmetry is the circumvention of the $\sqrt{3}$ reduction of the effective field that results from the weighted averaging nature of Floquet-based approaches\cite{Zhou2020}. 
In addition, another benefit of the perpendicular field configuration is the unique opportunity of using two groups of NVs simultaneously, which leads to a factor of two enhancement in the optical spin contrast. Finally, the sensitivity is further enhanced due to the larger magnetic moment difference in the perpendicular field encoding ($\Delta\mu\equiv\mu_{\ket{\tilde{B}}}-\mu_{\ket{\tilde{0}}}=1.15\mu_B$) compared to the on-axis encoding ($\mu_{\ket{0}}-\mu_{\ket{-1}}=\mu_B$), making the former intrinsically more sensitive to magnetic fields. The differences between different encoding for quantum sensing are summarized in Fig.~\ref{fig3}(a).

We then experimentally compare the sensing performance under three scenarios: the on-axis field configuration with and without Floquet engineered $\mathrm{SU}\left(2\right)$ symmetry, and the two-groups perpendicular field configuration with native $\mathrm{SU}\left(2\right)$ symmetry. We first fix the phase accumulation time to be $7.2~\mu$s in all three scenarios, and compare the optical contrast response as a function of the target AC field amplitude (Fig.~\ref{fig3}(b)). Here we see a slower oscillation in the Floquet engineered case (dark green), consistent with the expected $\sqrt{3}$ reduction of the effective field. We also observe a significantly larger spin contrast in the two-groups perpendicular field configuration (red), in addition to a slightly faster oscillation than the aligned field case without Floquet engineering (black). These observations are consistent with theoretical expectations, and demonstrate the advantages of our approach. We then compare the sensitivity scaling with phase accumulation time in Fig.~\ref{fig3}(c), where we see a $2.6\times$ (i.e. $8.3$~dB) enhancement of the sensitivity compared with DROID-60 in the on-axis field configuration, consistent with the expected breakdown into benefits and overheads (see Supplement). The nT-level volume normalized sensitivity (1.09~nT$~\mu\mathrm{m}^{3/2}\mathrm{Hz}^{-1/2}$), reported here, is among the best values reported so far in literature\cite{Arunkumar2023,Masuyama2018,Wolf2015,Zhou2023}.

\textit{Applications in many-body physics.}---Another application of our technique is to study many-body dynamics at intermediate to late timescales, which is previously inaccessible due to limited coherence. In particular, spin transport (Fig.~\ref{fig4}(a)) in dipolar systems has been an active research topic, due to the interplay of long range interactions
and strong positional disorder\cite{Braemer2022,Zhao2025} in such systems. Here we probe spin transport through the infinite temperature local auto-correlator $C_{\mathrm{Local}}^{\mathrm{ZZ}}\equiv\overline{\langle s_i^z\left(t\right)s_i^z\left(0\right)\rangle_{T=\infty}}$ (also known as local spin survival  probability in literature\cite{Zu2021,Peng2023}), whose decay profile encodes the nature of the underlying transport dynamics\cite{Zu2021,Peng2023,Stasiuk2025}. As shown in Fig.~\ref{fig4}(b), we measure $C_{\mathrm{Local}}^{\mathrm{ZZ}}$ with a modified disorder-order protocol\cite{Martin2023}, where the disorder winding step ($+h_is_i^z$, with $h_i$ being the on-site disorder) and the subsequent $\frac{\pi}{2}$-pulse (gray) create an initial state with random spin polarization in YZ-plane, the dephasing step ($\tau^\prime$) singles out Z-axis for the measurement, and the disorder unwinding step ($-h_is_i^z$) refocus the spins before global readout. For detailed derivation of this protocol, we refer readers to the Supplement.

Before applying this protocol to spin transport measurements, we first benchmark it under free evolution, where no MW pulses are applied in the dashed box ``$H$" in Fig.~\ref{fig4}(b). 
In this regime, the strong on-site disorder $\sum_i h_i s_z^i$ - which exceeds the dipolar interaction by two orders of magnitude (see Supplement) - dominates the dynamics and leads to frozen transport.
Despite the absence of transport, we still observe extrinsic decay of the measured signal (Fig.~\ref{fig4}(c)) originating from imperfect unwinding, as the disorder $h_i$ is not completely static between the winding and unwinding steps. 
In Fig.~\ref{fig4}(c), we see that such extrinsic decay is much slower in the two-groups perpendicular field configuration (red), compared to the on-axis field case (black). This indicates a highly static on-site disorder in the former, which, when combined with the extended coherence $JT_2$, enables spin transport measurements at intermediate to late times.

We then use this protocol to measure spin auto-correlations at the $\mathrm{SU}\left(2\right)$ point (Fig.~\ref{fig4}(d)), with on-site disorder decoupled via concatenated XY8 sequence\cite{Khodjasteh2005} (see Supplement). 
While in previous  experiments\cite{Martin2023} a stretched-exponential decay of $C_{\mathrm{Local}}^{\mathrm{ZZ}}\left(t\right)$ was observed at early times—which is a common phenomenon in disordered spin systems\cite{Davis2023,Martin2023,Signoles2021,Choi2017Depolaization}, such functional form is expected to transition to an algebraic (power-law) decay at late times, signaling the onset of emergent hydrodynamics\cite{Zu2021,Peng2023,Stasiuk2025} associated to the conservation of global spin $\sum_i s_i^z$.
By leveraging the extended coherence time enabled by the two-groups perpendicular field configuration, we revisit this problem. 
We first compare the measured $C_{\mathrm{Local}}^{\mathrm{ZZ}}\left(t\right)$ to a stretched exponential model by plotting it on a loglog-log scale (Fig.~\ref{fig4}(d) left axis, dark red), where a stretched exponential appears as a straight line. 
The data agrees well with this form at early times, but transitions to a markedly slower decay at late times as expected. 
We emphasize that such slowing-down is inaccessible in the on-axis field configuration\cite{Martin2023}, where the much shorter extrinsic decay timescale (dashed vertical line) obscures the intrinsic many-body dynamics. 
To further probe the nature of transport, we replot the same data on a log-log scale (right axis, pink), where power law decay manifests as a straight line and the power law exponent (i.e. slope) encodes the transport universality class\cite{Zu2021,Peng2023,Stasiuk2025}. Here we observe that the data approaches the diffusive scaling ($t^{-3/2}$, black) at late times, consistent with the onset of emergent spin diffusion.

\textit{Discussion and outlook.}---In this work we developed a method for direct tuning of the native dipolar interaction within a high density ensemble of NV centers. Our method is based on dressed-state qubit encoding under an external B-field perpendicular to the NV axis, in combination with a pulsed field that restores high quality readout in such configuration\cite{Gao2025}. We demonstrate improved coherence with this technique, and its unique applications in quantum sensing and quantum many-body physics at intermediate to late times.

Our work also opens up several new opportunities for future studies. In particular, the significant improvement of the dimensionless coherence parameter $JT_2$ enables the exploration of a wide range of exotic many-body phenomena—including emergent hydrodynamics\cite{Zu2021,Peng2023,Stasiuk2025} and finite temperature XY ferromagnetic ordering\cite{Block2024}—that happen at a previously inaccessible timescale. In addition, the long coherence achieved in this work, when combined with recently developed magnetic field gradient techniques\cite{Put2025}, paves the way towards the experimental realization of collective spin squeezing dynamics predicted for $\mathrm{SU}\left(2\right)$-interacting ensembles with a spin spiral initial state\cite{SpiralSqueezingTheory}.
Realization of such squeezing dynamics and its accompanying signal amplification\cite{Gao2025} would enable further improvements on quantum sensing, that has important applications on micro- and nano-scale imaging of condensed matter materials\cite{Casola2018} and biological structures\cite{Mohan2010,Choi2020Worm}. 
Furthermore, an interesting direction for future works is the integration of our method with high-fidelity spin readout techniques\cite{Arunkumar2023,Maier2025,Shields2015}, pushing the limits of magnetic sensitivity in nano-scale applications.
Finally, our method is expected to generalize beyond NV centers, and apply to any higher spin ($S>\frac{1}{2}$) system with $\pm m$ degeneracy and non-zero zero-field splitting (see Supplement).
\newline

We thank Nazli~U.~Koyluoglu and Andrew Maccabe for helpful discussions and James MacArthur for technical contributions. This work was supported
by  the National Science Foundation (grant number PHY-2012023), the Center for Ultracold Atoms (an NSF Physics Frontiers Center), Gordon and Betty Moore Foundation Grant No. 7797-01, the U.S. Department of Energy [DOE Quantum Systems Accelerator Center (Contract No.: DE-AC02-05CH11231) and BES grant No. DE-SC0019241], and the Army Research Office through the MURI program grant number W911NF-20-1-0136. We acknowledge the use of shared facilities of the UCSB Quantum Foundry through Q-AMASE-i program (NSF DMR-1906325), the UCSB MRSEC (NSF DMR 1720256), and the Quantum Structures Facility within the UCSB California NanoSystems Institute. 
A.B.J. acknowledge support from the NSF QLCI program through grant number OMA-2016245. 
L.B.H. acknowledges support from the NSF Graduate Research Fellowship Program (DGE 2139319) and the UCSB Quantum Foundry.

\bibliography{main}

\end{document}